\documentclass[prb,aps,twocolumn,showpacs,floats,amssymb,amsfonts]{revtex4}

\usepackage{graphicx}
\usepackage{bm}

\begin{document}

\title{
 Orbital ordering, Jahn-Teller distortion, and resonant x-ray scattering in 
 KCuF$_3$
}

\author{N. Binggeli}
\affiliation{Abdus Salam International Center for Theoretical Physics and
INFM DEMOCRITOS National Simulation Center, Strada Costiera 11, 34014 Trieste, Italy}

\author{M. Altarelli}
\affiliation{Sincrotrone Trieste, Area Science Park, 34012 Basovizza, Trieste,  
 and Abdus Salam International Center for Theoretical Physics,
         34014 Trieste, Italy  }

\date{\today}

\begin{abstract}

The orbital, lattice, and spin ordering phenomena in KCuF$_3$ are investigated
by means of LDA+U calculations, based on {\it ab initio} pseudopotentials.
We examine the Cu-$3d$ orbital ordering and the associated Jahn-Teller distortion in
several different  spin-ordered structures of KCuF$_3$.  The ground state is correctly 
predicted to be an A-type antiferromagnetic structure, and the calculated 
Jahn-Teller distortion agrees also  well with experiment. Concerning the
orbital ordering, we find that even for a highly ionic compound such as 
KCuF$_3$, the orbital-order parameter is significantly reduced with respect to 
its nominal value due to Cu($3d$)--F($2p$) hybridization.   We also calculate the
Cu K-edge resonant x-ray scattering spectra  for Bragg reflections associated with 
orbital order. Consistent with previous studies, we find that the resonant
signal is dominated by  the structural anisotropy in the distribution of the F
neighbors of the resonant Cu atom, and that the Cu-$3d$ orbital ordering has only
a minor influence on the spectra.  Our  LDA+U results, however,  also indicate that 
a change in the magnetic structure has a small  influence  on the Jahn-Teller 
distortion, and hence on the resonant  spectrum,  in the conventional  
(room-temperature) crystallographic structure of KCuF$_3$. This  may 
indicate that the large change observed experimentally  in the resonant signal 
near the N\'eel temperature is related to  a low-temperature structural 
transformation  in KCuF$_3$. 

\end{abstract}

\maketitle

\section{Introduction}

The pseudocubic perovskite KCuF$_3$ is a  magnetic insulator that has 
attracted significant  interest since the sixties, as a prototype material for 
orbital ordering, cooperative Jahn-Teller distortion, and low-dimensional  
magnetism.\cite{Kugel73,Goodenough63,Hutchings69}
KCuF$_3$ is also structurally related to  high-T$_c$
superconducting cuprates and colossal-magnetoresistance  manganites. 
In recent years, renewed attention has been given to the problem of 
orbital ordering and to the study of the interactions between orbital,
magnetic, and structural ordering in strongly correlated $3d$-transition-metal
compounds, stimulated in large part by the interest in colossal-magnetoresistance 
manganites.\cite{Oles00} In the latter  systems, the  interplay 
between orbital, charge, spin, and lattice degrees of freedom is known to play 
an  essential role in establishing the observed physical properties. Contrary to 
the case of charge, spin, and lattice  ordering, however, which can be 
investigated by conventional  x-ray, neutron, and electron diffraction
techniques, direct experimental observation 
of orbital ordering has been---and remains in most cases---a challenge. 

Recently, resonant elastic x-ray scattering (RXS) techniques have been 
developed to probe orbital ordering. RXS has been
applied to a number of orbitally-ordered  compounds, including various
manganites\cite{Murakami98,Zimmermann01} and
KCuF$_3$.\cite{Paolasini02,Caciuffo2002}
Except  for very recent measurements carried out on layered/doped
manganites at the Mn $L_{2,3}$
edges,\cite{Wilkins03a,Wilkins03b,Dhesi04,Thomas03} most of the
experiments performed to date have used excitations from the
transition-metal 1$s$ core level, i.e., K-edge excitations.
In the case of KCuF$_3$ and for related perovskite manganites such as
LaMnO$_3$, the K edge excitation is the only possible core-level
excitation for which resonant photons have a sufficiently large
wavevector  to access a Bragg vector of the targetted orbitally ordered
structure.  
The K-edge scattering process  in manganites and KCuF$_3$, however,  
derives from virtual electric dipole excitations to empty
transition-metal 4$p$ states. The resulting RXS amplitude is thus not
directly related to the  $3d$ states that exhibit orbital ordering, but to the  
anisotropy induced in the surrounding 4$p$ states. Although the  latter 
is due, in principle, to the orbital ordering and/or associated Jahn-Teller
distortion,  this indirect dependence complicates the interpretation of  
K-edge RXS data. In fact, it has been first proposed that the measured 
anisotropy is determined by the 3$d$ orbital polarization 
via onsite 3$d$--4$p$ Coulomb interaction.\cite{Ishihara98}
However, subsequent local-density-approximation (LDA) calculations and 
LDA plus onsite-Coulomb interaction (LDA+U) calculations, performed for a number
of manganites\cite{Elfimov99,Benfatto99,Benedetti01,DiMatteo03} and also for
KCuF$_3$,\cite{Caciuffo2002,Igarashi03} have provided rather strong evidence
that the K-edge scattering amplitude  in these systems is mostly sensitive to the 
Jahn-Teller distortion that accompanies the orbital ordering, rather than to 
the orbital ordering itself.

In KCuF$_3$, an interesting feature has been observed  concerning
the temperature dependence of the K-edge RXS intensity for orbital Bragg
peaks.\cite{Paolasini02,Caciuffo2002} The measured scattering intensity 
shows  a substantial and sudden increase (within 4-5 K), with
decreasing temperature,  when approaching  the N\'eel temperature.
A somewhat similar phenomenon has been observed also in LaMnO$_3$ at 
the K edge\cite{Murakami98} and in  La$_{0.5}$Sr$_{1.5}$MnO$_4$ at the 
$L_{2,3}$ edges,\cite{Wilkins03b} for
temperatures approaching the N\'eel temperature, but in these systems the
changes in the orbital scattering intensity
take place over a much wider temperature range, i.e., one to two orders of
magnitude wider. These phenomena are therefore not necessarily related, but
they are all very puzzling, because they take place in  a temperature region where 
both the orbital-order parameter and the Jahn-Teller distortion are expected 
to be saturated.  In all of these systems, the origin of the changes in the
scattering intensity is an unresolved issue.

For KCuF$_3$, two different interpretations have been proposed. 
On one hand, the change in RXS intensity has been ascribed to a change in the orbital 
order parameter, and interpreted as indicative of a  strong coupling between spin 
and orbital degrees of freedom.\cite{Paolasini02,Caciuffo2002} On the other hand, 
based on the results of  LDA + U calculations showing a dominant influence of the 
Jahn-Teller distortion on the RXS spectra, the jump in intensity has been attributed 
to an increase of the Jahn-Teller distortion in the low-temperature magnetic
phase.\cite{Igarashi03}
These interpretations are also not necessarily mutually exclusive;
a change in the magnetic structure may induce a change in the orbital-order
parameter, which in turn may modify the Jahn-Teller distortion. 
In principle, first-principle calculations could help discriminating 
between these various possibilities by providing  
quantitative information on the  influence of the magnetic structure on both, the
Jahn-Teller distortion and the orbital-order parameter. 
As yet, however, to our knowledge these issues have not been addressed by 
such calculations for  KCuF$_3$. 

In the present work, we use state-of-the-art LDA+U calculations,
based on {\it ab initio} pseudopotentials, to investigate the electronic, 
structural, and magnetic properties of KCuF$_3$. Our main targets are 
the orbital ordering and Jahn-Teller distortion, and we address the influence of 
the magnetic structure on these properties.  We also evaluate the Cu K-edge
RXS spectrum, and investigate the influence of the magnetic structure 
and  Jahn-Teller distortion on the simulated spectrum. 
Unlike other density-functional-related approaches, the LDA+U method
has not been extensively used to examine structural properties. In this study 
we therefore also first demonstrate the ability of LDA + U pseudopotential 
calculations to provide a realistic  description of the  ground-state 
properties of KCuF$_3$, when full structural  optimization is performed.  

\begin{figure}[t]
\includegraphics[width=8cm]{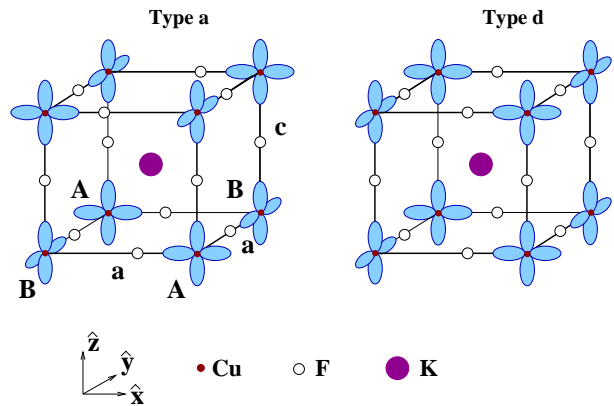}
\caption{\label{fig:cell}
(Color online) Schematic views of the atomic structure and  of  the  ordering of  the 
Cu-$3d$ hole orbitals in the a- (left) and d- (right) polytypes of KCuF$_3$. 
}
\end{figure}

KCuF$_3$ crystallizes in a perovskite structure which is tetragonally  
distorted   by  a collective Jahn-Teller distortion of
the CuF$_6$  octahedra.  Two different  polytypes
have been identified experimentally  at room temperature,
which differ in their atomic-plane stacking  along the $c$ axis
(Fig.~\ref{fig:cell}).\cite{Okazaki69}  
The Cu  ions in KCuF$_3$ are nominally in a Cu$^{2+}$ (d$^9$)  
configuration, with completely filled $3d$-$t_{2g}$ orbitals and  one
hole in one of the cubic degenerate $3d$-$e_g$ orbitals
($3d_{z^2-x^2}$, $3d_{3y^2-r^2}$).
The cubic degeneracy of the $e_g$ states is lifted  in the paramagnetic 
phase, well above room temperature,  by an orbital polarization and 
Jahn-Teller instability. 
The octahedra elongate along the $a$  or $b$ principal axes, with 
an alternate order on neighboring Cu A and B sites in the $ab$ plane
(see Fig.~\ref{fig:cell}). 
This distortion corresponds to an alternate distribution of 
$3d_{z^2-x^2}$ and $3d_{z^2-y^2}$ hole orbitals
on Cu A and B sites.\cite{Goodenough63}
Along the $c$ axis, the stacking of the planes may be 
antiferromagnetic-like (type $a$) or ferromagnetic-like (type $d$),
giving rise to the two different polytypes. 

Below the N\'eel temperature (38 K for polytype $a$,  22 K for
polytype $d$), the spins arrange in an A-type antiferromagnetic (AF) 
structure, with strong AF superexchange interaction between  
adjacent Cu atoms along the $c$ axis, and weak ferromagnetic (FM) 
superexchange coupling  between Cu nearest neighbors within the 
$ab$ plane.\cite{Hutchings69,Satija80}
This magnetic structure is consistent with the predictions of the 
Goodenough-Kanamori-Anderson\cite{Goodenough63} rules for the
superexchange interactions associated with the orbital-ordered
configurations displayed in  Fig.~\ref{fig:cell}. 
Such orbital configurations have been confirmed by previous LDA+U and 
Hartree-Fock calculations for
KCuF$_3$.\cite{Caciuffo2002,Igarashi03,Liechtenstein95,Dovesi95,Anisimov02} 
Previous calculations have also shown that the LDA  approach 
fails to correctly predict the stable orbitally-ordered AF insulating  
structure of  KCuF$_3$, while
the LDA+U corrects this feature.\cite{Liechtenstein95} 

In this paper, we concentrate on the $a$-type structure of KCuF$_3$, which has been
the  focus of the experimental RXS studies.\cite{Paolasini02,Caciuffo2002}
The work is organized as follows: In Section~\ref{sec:Method},
we briefly present the calculation method. In Section~\ref{sec:Lattice}, 
we address the structural properties of the low-temperature 
A-AF structure; the corresponding orbital ordering and electronic properties are 
examined in Section~\ref{sec:Orbital}. In Section~\ref{sec:Spin}, we investigate 
the energetics and   
superexchange couplings of several different (meta)-stable spin structures of 
KCuF$_3$. The influence of the magnetic structure on the orbital ordering and 
Jahn-Teller distortion is addressed in  Section~\ref{sec:Influence}.
In Section~\ref{sec:RXS}, we present our results for the RXS  spectra, 
and we summarize our conclusions in Section~\ref{sec:Summary}.

\section{Methodology}\label{sec:Method}

The calculations are performed within the LDA + U scheme, using
the pseudopotential plane-wave method. We employ 
the  rotational invariant\cite{Liechtenstein95}  LDA + U
energy functional:\cite{Cococcioni,Sawada97}
\begin{eqnarray}\label{eq:functional}
E^{\text{LDA+U}} [\rho^{\uparrow},\rho^{\downarrow},
\{ n^{I,\uparrow}_{\lambda}\},\{ n^{I,\downarrow}_{\lambda}\}] =
 E^{\text{LDA}} [\rho^{\uparrow},\rho^{\downarrow}] &+& \nonumber \\
\frac{U-J}{2} \sum_{I,\lambda,\sigma} n^{I,\sigma}_{\lambda}
 \left( 1 - n^{I,\sigma}_{\lambda} \right), &&
\end{eqnarray}
where $E^{\text{LDA}} [\rho^{\uparrow},\rho^{\downarrow}]$ is
the standard LDA energy functional, $\rho^{\sigma}$ is the
electron spin density with spin polarization $\sigma$, $U$ 
($J$) is the onsite Coulomb (exchange) interaction
parameter for the localized electrons---the
Cu-$3d$ electrons, and $n^{I,\sigma}_{\lambda}$ are the
occupation numbers of the  localized orbitals. I and
$\lambda$ are indexes for atomic sites and  orbitals, respectively.
The occupation numbers  $n^{I,\sigma}_{\lambda}$ are the
eigenvalues of the  density matrices:
\begin{equation}\label{eq:densmat}
n^{I,\sigma}_{m,m'} = \sum_{\bm{k},i} f^{\sigma}_{\bm{k},i}
\langle \varphi^{I,\sigma}_m | \psi^{\sigma}_{\bm{k},i} \rangle
\langle \psi^{\sigma}_{\bm{k},i}| \varphi^{I,\sigma}_{m'} \rangle,
\end{equation}
where $\varphi^{I,\sigma}_m(\bm{r})$ are the rotational-degenerate 
atomic orbitals of the localized-$d$-electron shell, $m$ stands for the 
angular quantum number,   
$\psi^{\sigma}_{\bm{k},i}(\bm{r})$ are the crystal Bloch states,
and $f^{\sigma}_{\bm{k},i}$ are the corresponding occupation
numbers. 

In our pseudopotential calculations, the crystal Bloch states, 
$\psi^{\sigma}_{\bm{k},i}(\bm{r})$, are
represented by Bloch pseudo wavefunctions, and the matrix elements
$\langle \varphi^{I,\sigma}_m | \psi^{\sigma}_{\bm{k},i} \rangle $, in 
Eq.~(\ref{eq:densmat}), are evaluated by projecting the crystal pseudo
wavefunctions on atomic pseudo  wavefunctions.\cite{Note_proj}
We employ the Perdew-Zunger parametrization of
the LDA exchange-correlation potential.\cite{PZ} For the effective
onsite Coulomb interaction parameter, $U_{\rm eff} = U - J$,
we use a value of 6.6~eV.  This value corresponds to the calculated values
$U = 7.5$~eV and $J = 0.9$~eV obtained by constrained LDA
computations for KCuF$_3$ in Ref.~\onlinecite{Liechtenstein95}. 

We employ a Vanderbilt\cite{Vanderbilt}  ultrasoft pseudopotential  for Cu and
norm-conserving Troullier-Martins\cite{TM}  pseudopotentials  for K and F.  
Some of the calculations  have also been performed using a  Troullier-Martins  
pseudopotential for Cu,\cite{TM} for comparison. 
The pseudopotentials we use are scalar relativistic. 
The  Vanderbilt pseudopotential has been generated in the non-magnetic Cu 
$3d^{9.5} 4s^{0.5} 4p^{1}$ atomic configuration, using for the  core-cutoff radii: 
$r_{\rm s} = 2.1$, $r_{\rm p} =  2.2$, and $r_{\rm d} = 2.0$ a.u.. 
The Troullier-Martins  pseudopotentials have been generated
in the ground-state  configuration of the non-spin-polarized 
atom,  with the cutoff radii (in a.u.):
$r_{\rm 4s} = 3.50$, $r_{\rm 4p} = r_{\rm 3d} = 3.75$ for K and 
$r_{\rm 2s} = r_{\rm 2p} = 1.40$ for F. 
The Troullier-Martins pseudopotentials have been
cast into the Kleinmann-Bylander\cite{KB}   non-local form using the
$s$-component as local part.  For Cu and K, we have included the non-linear core 
correction\cite{nlcc} to the exchange-correlation potential  to account for the 
overlap  between the valence and core charge;  for fluorine this is not necessary, 
since we treat the F $2s$ semicore electrons as valence electrons.

All calculations are carried out  in a 20-atom  tetragonal unit cell. 
We use a kinetic-energy cutoff of 80 Ry for the plane-wave expansion of the 
electronic states. The  integrations in reciprocal space are
performed using a (4,4,2) Monkhorst-Pack\cite{MP} k-point grid for all
structures, except the non-magnetic structure of KCuF$_3$ which is metallic.
For the latter structure we employ a (6,6,4) Monkhorst-Pack grid together with
a Gaussian electronic-level broadening scheme,\cite{Fu83} with a
full width at half maximum of 0.01~Ry.
With  the above parameters, the estimated convergence on the relative energies
of the various magnetic structures considered in this work is $\sim1$~meV 
per formula unit. 

Concerning the pseudopotentials,  we have performed some comparative studies  
using the Troullier-Martins and Vanderbilt pseudopotential for Cu.  
They indicate negligible differences, in general, for results concerning 
the electronic-structure properties.   
For the structural properties,  the Vanderbilt pseudopotential yields a larger  
($\sim15$\,\% larger) Jahn-Teller  distortion, and a smaller equilibrium volume
(2\,\% smaller) than the  Troullier-Martins pseudopotential. This is related to a 
higher degree of localization within the core region of the Cu-$3d$ pseudo-charge
density obtained with the Vanderbilt compared to the Troullier-Martins 
pseudopotential. The structural trends,  however,  and in particular the 
dependence on the magnetic state,  are 
the same with the two types of pseudopotentials. 
The effect of the K $3p$ and $3s$ semicore electrons has 
also been investigated.  Potassium is known to have  semicore states which 
are rather polarizable.  We have therefore performed  test calculations treating  
the K $3p$ and $3s$  electrons as valence electrons. They show that the  
inclusion of the K semicore electrons  has virtually no effect on the  electronic 
properties and changes  the equilibrium values of the structural parameters by less
than one percent. 

\section{Structural properties}\label{sec:Lattice}

In Table~\ref{tab:equilibrium}, we show our  results for the
equilibrium lattice constant, $a$, tetragonal lattice ratio $c/a$, fluorine
coordinate $x_F$, and  bulk modulus $B_0$ of the low-temperature 
A-AF structure of KCuF$_3$.  The theoretical values of the structural 
parameters are compared to experimental values at low  temperature
---when available---and at room temperature. 
The theoretical equilibrium lattice constant $a$ is about 2\,\% smaller
than the measured low-temperature value. This is comparable to 
LDA results for  related perovskite materials, such as the ferroelectric 
oxides studied in Ref.~\onlinecite{KingSmith94}, for which the LDA
underestimation of the lattice constant is typically  $\sim1$ to 2\,\%.

\begin{table}[t]
\begin{ruledtabular}
\begin{tabular}{lcccc}
 & a  (\AA) &   c/a  & x$_F$ & B$_0$ (GPa)  \\ \hline
LDA + U      & 4.03  &  0.955 & 0.2320 & 75  \\
Expt$^{a}$ (10 K)   & 4.126    &  0.9486   &    &     \\
Expt$^{b}$ (298 K)  & 4.141    &  0.9476   &  0.2280 &     \\
\end{tabular}
\end{ruledtabular}
\caption{\label{tab:equilibrium} Calculated values of the
equilibrium lattice constant, $a$, tetragonal lattice ratio, c/a
internal structural parameter, x$_F$, and
bulk modulus, B$_0$, of KCuF$_3$.
The experimental data are from Ref. \onlinecite{Satija80} (a)
and from Ref. \onlinecite{Buttner89} (b).
}
\end{table}

Our LDA+U values for $c/a$ and  $x_F$  agree
well with experiment, i.e., to within 1 and 2\,\%, respectively.
The calculated relaxed bond lengths for the three inequivalent Cu--F bonds
along the $x,y$, and $z$ directions (at a Cu A site)
 are 1.87, 2.16 and 1.92 \AA, respectively, versus 1.88,  2.25
and 1.96~\AA\  in the experiment.\cite{Buttner89}
The  in-plane quadrupolar distortion
$\Delta X = (0.25 - x_F) 2 a $, associated with the Jahn-Teller
distortion of the CuF$_6$ octahedra, is 3.6\,\% of $a$, compared to the
experimental value of 4.4\,\%.\cite{Buttner89}
Hartree-Fock calculations\cite{Dovesi95} performed for the $d$-type 
structure yield a distortion $\Delta X/a$ of 2.6\,\%;
the same Hartree-Fock calculations corrected {\it a posteriori}
using a gradient-corrected correlation functional give a
distortion of 3.4\,\%, closer to our calculated value.
Earlier LDA + U calculations, based on the full-potential
linear-muffin-tin-orbital  method and performed using the
experimental values of lattice constants, reported
a smaller quadrupolar distortion of 2.5\,\% for
the $d$-type structure.\cite{Liechtenstein95}
To our knowledge, no experimental value is available for the
bulk modulus of KCuF$_3$. Our value of 75 GPa, however, is in
reasonable agreement with the $B_0$ value of 84 GPa derived
from semiempirical pair-interaction-model calculations for 
the elastic constants of KCuF$_3$.\cite{Nikiforov96}

\begin{figure}[t]
\includegraphics[width=7.8cm]{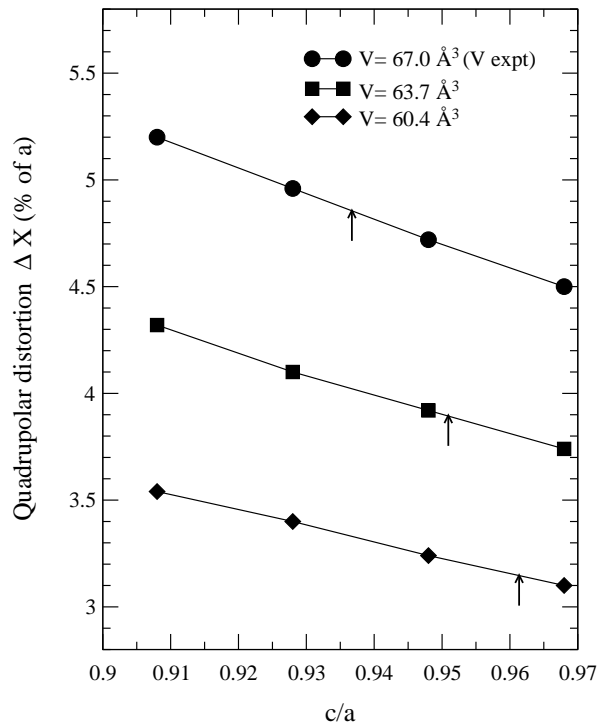}
\caption{\label{fig:quad}
Calculated quadrupolar distortion $\Delta X = (0.25 - x_F) 2 a $
as a function of the cell-shape ratio $c/a$ for different
volumes  of KCuF$_3$. The arrows indicate the equilibrium values
under hydrostatic conditions.
 }
\end{figure}

The structural optimization needed to determine the
parameter values in Table~\ref{tab:equilibrium} was carried out
considering several different unit-cell volumes, and for each volume
several different $c/a$ ratios, and by fully relaxing in each case the
F internal coordinates. As a by-product, we have therefore  access
to the dependence of the quadrupolar distortion on tetragonal 
strain and also to the hydrostatic-pressure dependence of $c/a$ and $x_F$.
In Fig.~\ref{fig:quad}, we display the calculated equilibrium quadrupolar
distortion $\Delta X$ as a function of the $c/a$ ratio,  for
three different volumes (including the experimental one).
Consistent with semiempirical model descriptions,\cite{Goodenough63}
an increase in $\Delta X$ decreases $c/a$; the results in
Fig.~\ref{fig:quad} indicate that a 10\,\% increase in 
$\Delta X$ decrease $c/a$ by $\sim5$\,\%.
In Table~\ref{tab:structure_p}, we also reported the equilibrium
values of $x_F$ and $c/a$ at different volumes.
Decreasing the volume, decreases the quadrupolar distortion, and
increases the $c/a$ ratio. The latter trend is not inconsistent with
the  increase in the experimental $c/a$ ratio observed  with decreasing
temperature (and volume) in  Table~\ref{tab:equilibrium}.

In the following sections, we  report our results for the electronic
properties, magnetic structure, and RXS spectra of KCuF$_3$ evaluated at
the reference experimental values of the structural parameters. We note that
using the theoretical values of the structural parameters would not change
significantly the results, i.e., produce negligible changes in the
orbital-order parameters, variations of $\sim0.01 \mu_B$ in the magnetic
moments, and changes of the order of 0.1~eV in the relative energies of
the electronic states.

\begin{table}[t]
\begin{ruledtabular}
\begin{tabular}{lcc}
      V (\AA$^3$) &  c/a & x$_F$ \\  \hline
67.03 & 0.937& 0.2257\\
63.68 & 0.951& 0.2305\\
60.39 & 0.961& 0.2343\\
57.24 & 0.968 & 0.2371\\
 \end{tabular}
\end{ruledtabular}
\caption{\label{tab:structure_p} Calculated equilibrium values of
the tetragonal lattice ratio c/a and internal structural parameter
$x_F$ of KCuF$_3$ at different volumes.
}
\end{table}

\section{Electronic structure and orbital ordering}\label{sec:Orbital}

\begin{figure}[b]
\includegraphics[width=7.2cm]{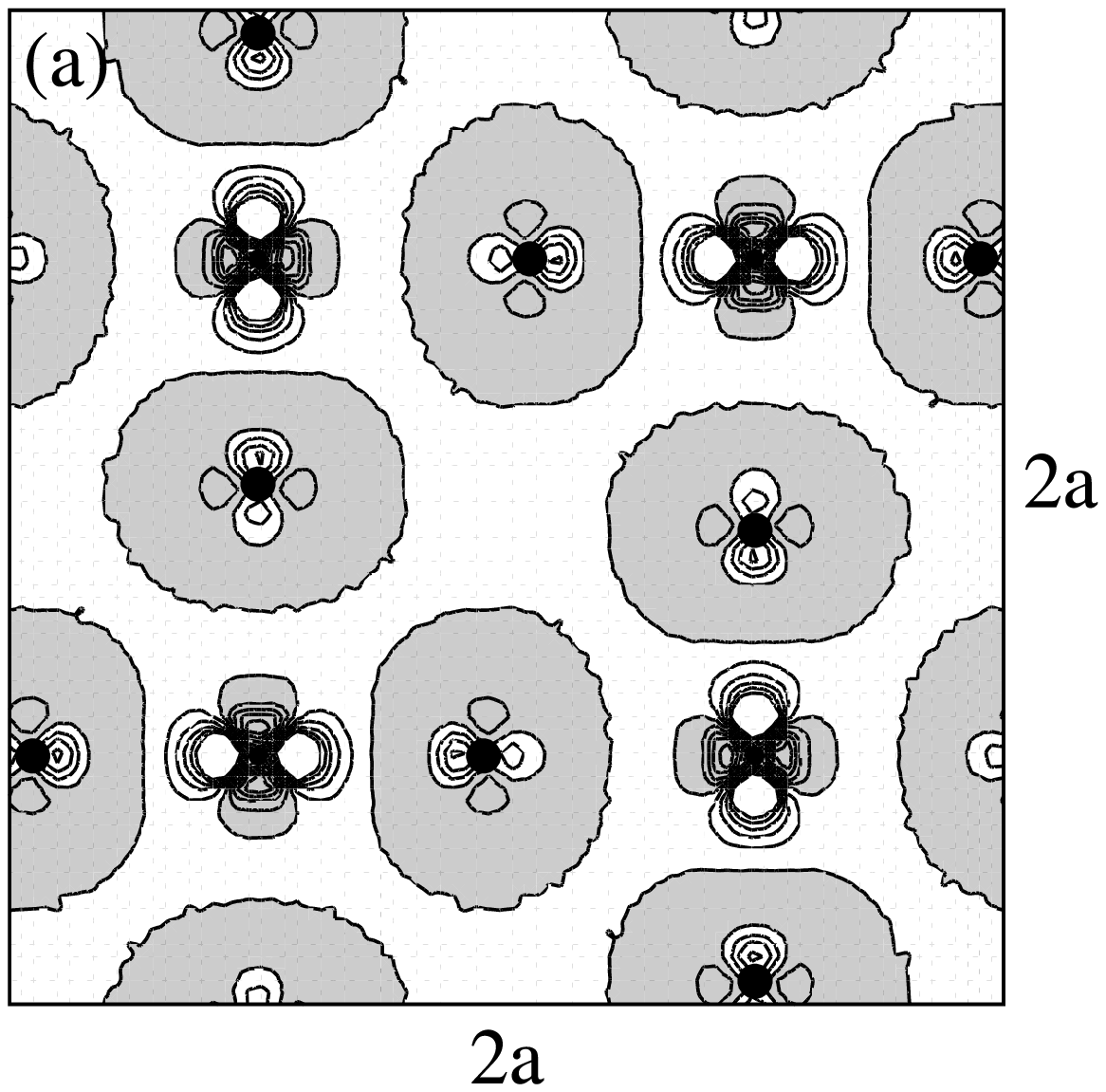}

\vspace*{0.2cm}

\includegraphics[width=7.2cm]{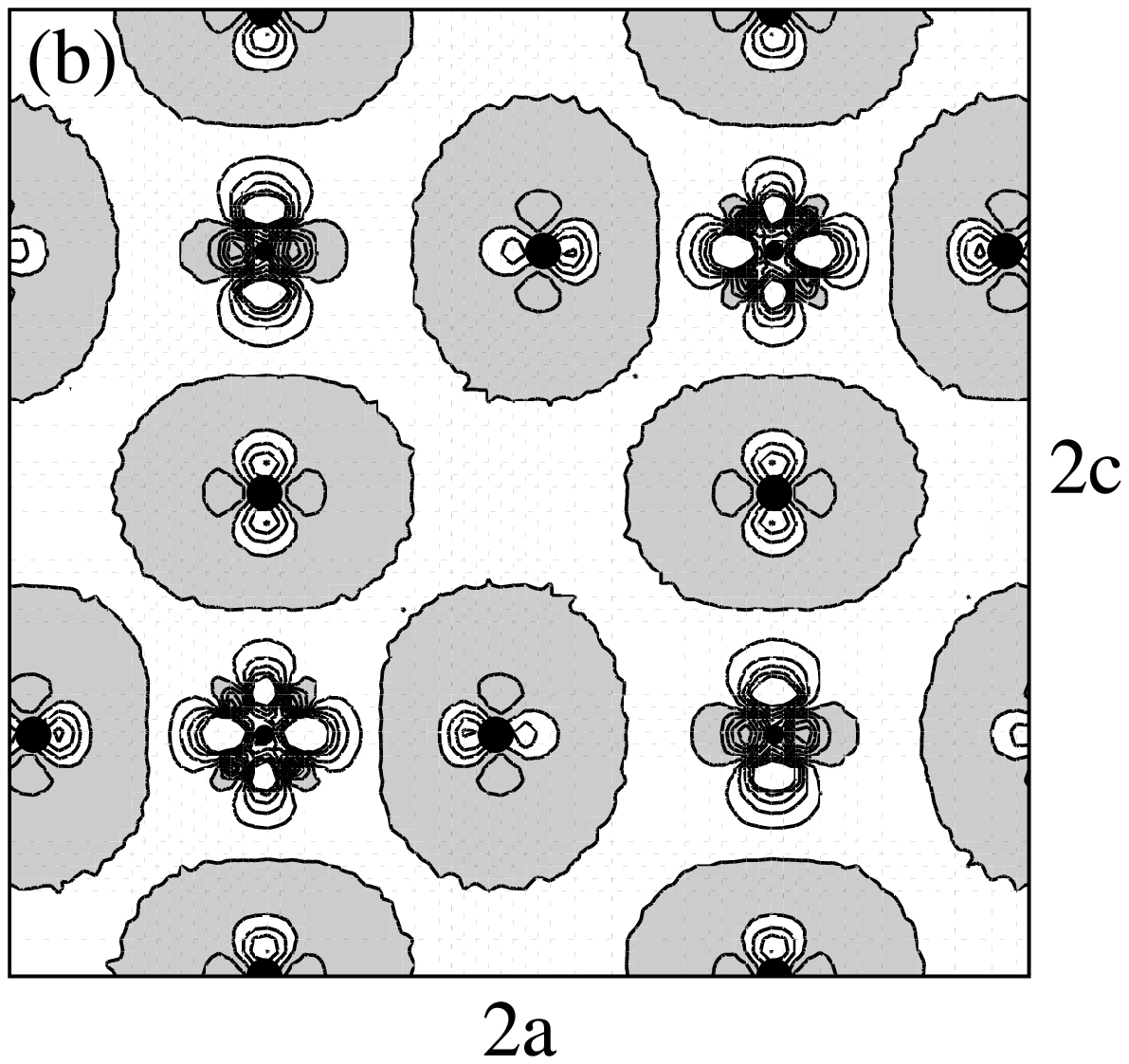}
\vspace*{0.0cm}

\caption{\label{fig:defmap}
Difference between the crystal electronic charge density
and the superposition of spherical atomic charge densities of 
neutral atoms in
a (001) (panel a) and a (010) (panel b) CuF atomic plane of
KCuF$_3$.  The positions of the Cu (F) atoms
are indicated by small (large) dark disks.  Regions of electron 
accumulation (depletion) are shown in gray (white). Contour
lines are separated by 0.03 $|e|/{\rm bohr}^3$; negative electron 
densities are truncated below -0.15 $|e|/{\rm bohr}^3$ (near the 
Cu atoms).
}
\end{figure}

In Fig.~\ref{fig:defmap}, we present the KCuF$_3$  differential charge
density for the A-AF structure, i.e., the difference between the crystal 
electronic charge density and the superposition of spherical atomic
 charge densities 
of neutral atoms.
The differential density provides information on the electronic
charge rearrangements involved in the crystal formation, including
charge transfers  associated with bond formation and 
orbital ordering. 
The differential charge is displayed both in a  CuF (001) 
 [Fig.~\ref{fig:defmap}(a)] and  in a  CuF (010) 
[Fig.~\ref{fig:defmap}(b)]  atomic plane of KCuF$_3$.

The differential density is clearly partitioned into
F-related/Cu(K)-related electron accumulation/depletion regions,
consistent with the ionic nature of KCuF$_3$.
The Cu-$3d$ orbital ordering shows up quite strikingly in
the maps, i.e., in the alternate sequence of  depleted
$d_{z^2-x^2}$- and $d_{z^2-y^2}$-like orbital densities on
neighbouring Cu A and B sites in Figs.~\ref{fig:defmap}(a) and (b).
We notice a slight  asymmetry in the $d_{z^2-x^2}$-like
($d_{z^2-y^2}$-like) orbital density related to the tetragonal
distortion of the lattice, i.e., stronger depletions  for the lobes
along the $x$ (respectively $y$) axis  than for the lobes along the
$z$  axis.
We also observe a local increase   in  the  electronic density
of some of the nominally ``filled''  Cu-$3d$ orbitals---in particular
the $d_{3y^2-r^2}$ ($d_{3x^2-r^2}$) orbitals---suggesting
a slight contraction of the $3d$ states of the  Cu ion in the
crystal with respect to those of the isolated atom. 
We note that the relaxation of the  three inequivalent
Cu--F bonds  follows the trend of decreasing  bond length with
increasing  Cu-$3d$ localized-hole charge on the bond, 
as expected from electrostatic interaction with the F anions. 


\begin{figure}[t]
\includegraphics[width=8cm]{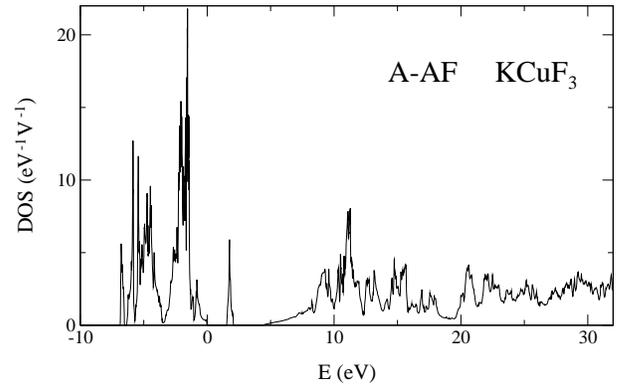}
\caption{\label{fig:dos}
Electronic density of states of A-AF KCuF$_3$. The zero of
energy   corresponds to the valence band maximum.
 }
\end{figure}

The differential maps  indicate a significant interaction
(hybridization) between Cu-$3d$ $e_g$ and F-2$p_\sigma$ states 
and a  related non-negligible delocalization of the Cu hole on the
nearest-neighbor F 2$p_{\sigma}$ states. These features, 
together with the deviation from purely 
$d_{z^2-x^2}$ ($d_{z^2-y^2}$) hole orbitals,  suggest a
somewhat incomplete $z^2-x^2/z^2-y^2$
orbital ordering.  To be more quantitative in this  
statement, we have evaluated, from the density matrix, the
total number of $3d$ holes per Cu site,
$n_{3d}^{hole} = 10 - \sum_{\sigma,\lambda} n^{\sigma}_{\lambda}$,
and the expectation value of the number of holes in the
$d_{z^2-x^2}$ orbital on an  A site (or equivalently,  
$d_{z^2-y^2}$ orbital on a B site), 
$\bar{n}^{A,hole}_{z^2-x^2} = 2 -
\sum_{\sigma} n^{A,\sigma}_{z^2-x^2,z^2-x^2}$.
We note that the occupation
numbers we obtain from the density matrix for the $t_{2g}$ orbitals
are identically 1.00 per spin, so that  the total  number of $3d$
holes is identical to the number of $e_g$ holes in our calculations. 
The calculated values  are: $n_{3d}^{hole} =
n_{e_g}^{hole} = 0.63$ and $\bar{n}^{A,hole}_{z^2-x^2} =
0.61$. Hence, the  degree of  $z^2-x^2/z^2-y^2$
orbital ordering, measured by  $\bar{n}^{A,hole}_{z^2-x^2}$,
is significantly reduced (by $39$ \%) with respect
to its nominal value of one. Most of this reduction, however,
can  be accounted for by the decrease in the number of $e_g$ hole, 
$n_{e_g}^{hole}$, relative to one ($37$ \%),  and derives  thus from the
delocalization of the $3d$ hole (due to hybridization) on neighboring  
F orbitals. The $d_{z^2-x^2}$ ($d_{z^2-y^2}$) character of the
ordered orbitals on the A (B) sites,  measured by the ratio
$ \bar{n}^{A,hole}_{z^2-x^2} / n_{e_g}^{hole} $,
is instead nearly complete
[i.e., 97\,\%  of the $d_{z^2-x^2}$ ($d_{z^2-y^2}$) type].

\begin{figure}[t]
\hspace*{0.0cm}\includegraphics[width=8.2cm]{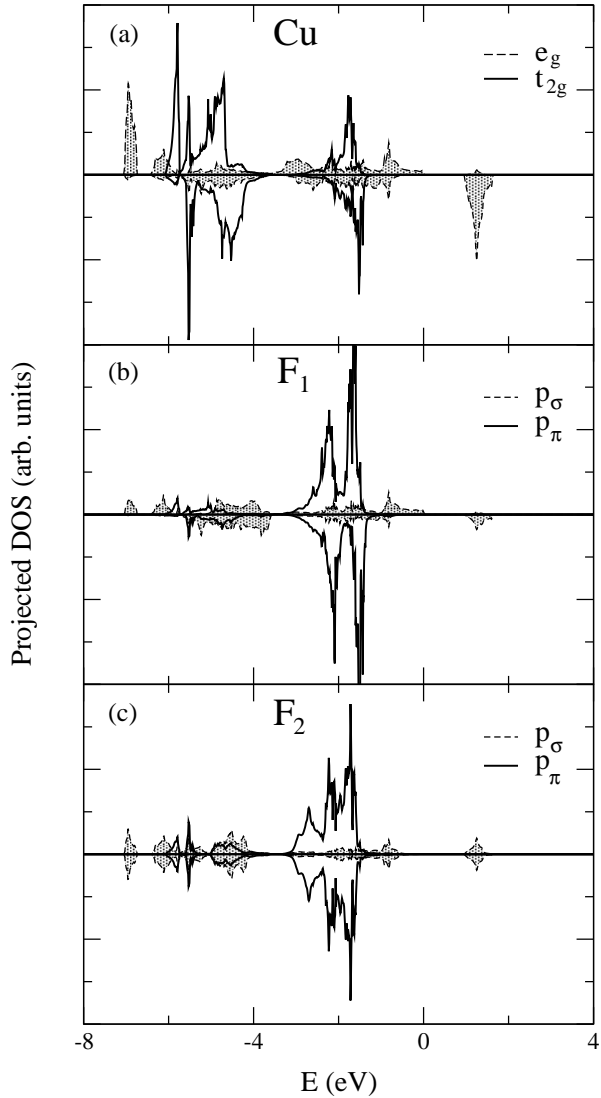}
\caption{\label{fig:partialdos}
Projected Cu-$3d$ and F-$2p$ atomic densities of states in KCuF$_3$.
Density of states projected onto: (a) the Cu $3d$ $e_g$
(dashed line) and $t_{2g}$  (solid line) orbitals,
(b) the F$_1$, and (c) the F$_2$ $2p_{\sigma}$ (dashed line)
and  $2p_{\pi}$  (solid line)  orbitals. In each panel, the upper 
half corresponds to the majority spin states and the lower half to the 
minority spin states.
 }
\end{figure}

The calculated electronic density of states (DOS)  of A-AF KCuF$_3$
is shown in Fig.~\ref{fig:dos}. 
KCuF$_3$ is  an insulator
in our calculations with a gap of 1.54 eV. The valence DOS of KCuF$_3$ 
exhibits two main structures: a   low-energy structure  
(-7 to - 3.5 eV), which   includes mostly Cu-$3d$ states, with an 
admixture of F-$2p$ states (mainly $2p_{\sigma}$ states), and a
high-energy structure (-3.5 to 0 eV), which is dominated by
F-$2p$ states, and  includes also a non-negligible contribution from
Cu-$3d$ states. The corresponding projected Cu-$3d$  and F-$2p$
atomic DOS's are displayed in Fig.~\ref{fig:partialdos}.\cite{Note_pdos}
The  Cu-$3d$ local atomic DOS has been split into contributions 
from the  $t_{2g}$ and $e_g$ states  [Fig.~\ref{fig:partialdos} (a)] 
and the F-$2p$ local DOS's  into contributions from the $p_{\sigma}$ 
and $p_{\pi}$ orbitals on each of the two
inequivalent F sites [Figs.~\ref{fig:partialdos} (b) and (c)].
The  sharp lowest-energy conduction-band feature, in Fig.~\ref{fig:dos},
clearly derives from the ordered Cu-$e_g$ empty orbitals, that have
the minority spin on each Cu site [see Fig.~\ref{fig:partialdos} (a)].
Within the LDA+U scheme, the lowest-energy conduction-band
feature may be viewed as the upper Hubbard band  associated with the 
localized $e_g$ orbitals illustrated in Fig.~\ref{fig:cell}, 
the corresponding lower Hubbard  band being the sharp 
lowest-energy valence-band feature in Fig.~\ref{fig:partialdos} (a).
The existence of a significant hybridization between Cu-$e_g$ and
F-$p_{\sigma}$ states,  in Fig.~\ref{fig:partialdos}, is apparent 
from the  energy degeneracies  and analogous dispersions of the 
related  features. Inspection of Fig.~\ref{fig:partialdos}
also reveals a non-negligible interaction between
Cu-$t_{2g}$ and F-$p_{\pi}$  states. 

\begin{figure}[t]
\includegraphics[width=8.2cm]{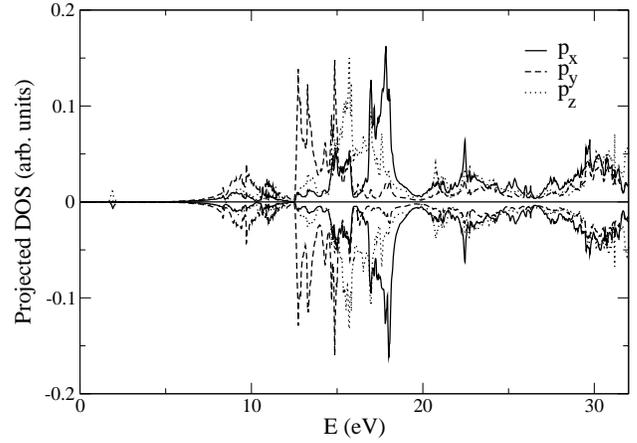}
\caption{\label{fig:pdos}
Density of states projected onto the Cu  $4p_x$ (solid line),
$4p_y$ (dashed line), and $4p_z$ (dotted line)
orbitals at a Cu A site.
 }
\end{figure}

At higher energy, in  Fig.~\ref{fig:dos}, one finds a second
conduction-band edge (at about 5 eV) associated with K-$4s$ states,
with also a contribution from Cu-$4s$ states. The two
sharp features located at $\sim9$ and 11~eV, in  Fig.~\ref{fig:dos},
originate from  K-$3d$ states.
We note that the Cu-$4p$ states, which are the   intermediate
states in the  Cu K-edge RXS process, contribute to the DOS over
a large energy region that ranges from about 8 eV to 32 eV (at
least). The corresponding Cu $4p_x$, $4p_y$, and
$4p_z$ projected densities of states are represented in 
Fig.~\ref{fig:pdos} (for a Cu A site).
The dominant features of the Cu $4p_x$ and $4p_y$ densities of 
states exhibit a splitting of about 6~eV
produced by the F quadrupolar distortion $\Delta X$.
The inward/outward relaxation of the nearest-neighbor F anions
along the $x$/$y$ axis tends to shift the $4p_x$/$4p_y$ states at 
higher/lower energy with respect to the $4p_z$ states.
It should be noted, concerning the high-energy unoccupied states 
occurring above 5 eV, that---unlike the lower-energy unoccupied
Cu-$d_{z^2-x^2}$ states associated with the sharp  
feature---these states are  totally  insensitive to the effective 
Hubbard term,
$U_{\rm eff}$,  in  our LDA+U calculations. These states are thus
expected to suffer from the usual LDA underestimation
of the gap (i.e., to be shifted to lower energy with respect
quasiparticle energies obtained in GW  calculations).

\section{Magnetic structures}\label{sec:Spin}

\begin{figure}[t]
\includegraphics[width=8cm]{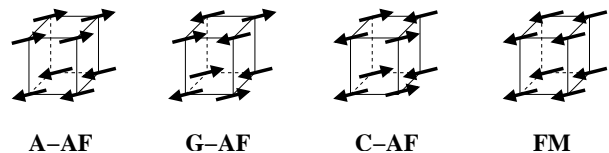}
\caption{\label{fig:mag}
Magnetic structures considered in this work for  KCuF$_3$.
 }
\end{figure}

The Goodenough-Kanamori-Anderson (GKA) rules predict,\cite{Goodenough63} 
for the half-filled-$d_{z^2-x^2}$/$d_{z^2-y^2}$ orbital
 configurations displayed in Fig.~\ref{fig:cell}, that the
favorable Cu superexchange interactions should be antiferromagnetic and
strong along the $c$ axis, and weakly ferromagnetic within the $ab$ plane.
In order to probe the magnetic interactions in our system, and
also to address the influence of the magnetic order on the electronic and 
structural properties (in the next section),  we have considered several
different possible (meta)-stable spin configurations for KCuF$_3$,
illustrated in Fig.~\ref{fig:mag}. In addition to the A-AF structure,
we have considered the AF G- and C-type configurations, the
FM state, and also the non-magnetic (NM) configuration.

In Table~\ref{tab:E_mag}, we give the relative energies of the various spin 
structures together with the corresponding values of the  ordered 
Cu spin moment $m_s$---computed as the integrated absolute value of the
magnetic moment per Cu atom.
In agreement with experiment, and consistent with the GKA rules, the
A-AF structure is the most stable configuration.
The second-most
stable configuration is the G-AF  spin structure, where the AF order
is preserved within the Cu chains along the $c$ axis, but FM
order has been replaced by AF order within the $ab$ plane. The structures
having, instead, FM order within the Cu $c$-chains (C-AF and FM in
Fig.~\ref{fig:mag})  are energetically  much less competitive.
For the latter structures, we note that AF spin order is more
favorable than FM spin order  within the $ab$ plane
(E[C-AF] $<$ E[FM]). This is contrary to the situation observed 
for the structures having  AF spin order within the chains, where  
FM order is preferred within the $ab$ plane (E[A-AF] $<$ E[G-AF]).
With the exception of the E[C-AF] $<$ E[FM] case, the energy
ordering we find is consistent with the trend expected from the
GKA rules for half-filled orbitals. Concerning the E[C-AF] $<$ E[FM]
anomaly, we note that in our system the hole orbital occupation differs
from the half-filled situation (see also next section), and
for the weak superexchange interaction within the $ab$
plane, such a deviation can change the sign of the superexchange
coupling.

The energy of  the NM structure, in Table~\ref{tab:E_mag},
is  by far ($0.2-0.3$~eV) the largest of all.  It should be noted, however,
that a large fraction of the energy difference between the  NM structure
and the stable A-AF phase can be accounted for by the  energy change
of in isolated Cu$^{2+}$ ion when its configuration is switched  from
non-spin polarized to spin polarized [i.e., $+0.24$~eV  in our LDA atomic
calculations for a spin moment of 0.95 $\mu_B$
(see Table~\ref{tab:E_mag})].
For the paramagnetic phase thus, with disordered  spins on the
different Cu sites,  a reasonable energy estimate  would be:
E[NM] $- 0.24$ eV $ \approx 0.06$~eV, which is
intermediate between the energy of FM and A-AF phase.

\begin{table}[b]
\begin{ruledtabular}
\begin{tabular}{lcc}
 & E (meV)&  m$_{\rm s}$ ($\mu_B$) \\ \hline
A-AF &  0  & 0.91\\
G-AF &  10 & 0.90\\
C-AF &  58 & 0.98\\
FM  &  70  & 1.03\\
NM & 297 & 0 \\
\end{tabular}
\end{ruledtabular}
\caption{\label{tab:E_mag} Calculated total energies and ordered Cu
spin moments in the various magnetic structures of KCuF$_3$
considered in this work.
Energies are given per formula units and measured relative to
the energy of the A-AF structure.
}
\end{table}

We note that the relative stability of the  different structures,
in Table~\ref{tab:E_mag}, is reflected in their   bandgaps.
We find that the bandgap changes from   1.54 eV,
for the A-AF structure, to 1.47, 1.00, 0.72, and 0 eV for the G-AF,
C-AF, FM, and NM structures, respectively.

In Table~\ref{tab:E_mag},  the  ordered moment $m_s$
of a Cu with a given spin $\sigma$  increases, in the
different spin structures,  with increasing number of Cu nearest neighbours
along the $c$ axis having the same spin $\sigma$.
We also observe an increase in the moment, but significantly  smaller,
with increasing number of  nearest neighbours having the same spin
within the $ab$ plane. These trends are consistent with the expected
Cu--F--Cu spin-density overlaps associated with the orbital arrangements
in Fig.~\ref{fig:cell}. The experimental value of the Cu ordered
spin moment  measured by neutron diffraction in the A-AF
structure at low temperature (4~K) is $\sim 0.49 \mu_B$,\cite{Hutchings69}  
which is significantly smaller than  our value of $0.91 \mu_B$. 
The small experimental   $m_s$ in KCuF$_3$, however, is generally
attributed to zero-point fluctuations in the spin direction (beyond
our calculations), which lower the expectation value of the onsite
spin moment.  This effect is especially large in KCuF$_3$, as a result
of the quasi-one dimensional nature of its magnetic structure.

Based on the spin-polarized   energies given in
Table~\ref{tab:E_mag}, one may derive some rough estimates for
the Cu superexchange coupling constants along the $c$ axis ($J_c$) 
and within the $ab$ plane ($J_a$). Using an Ising model---with
interactions $2 J_{c(a)} {\rm S_z^{(A)}} {\rm S_z^{(B)}}$
between spins ${\rm S } = \frac{1}{2}$ on adjacent A and
B sites along the $c$ axis (within the $ab$ plane)---to map our
calculated  magnetic energies, we  obtain for
$J_c$  values of $+70$ and $+48$~meV from the energy differences
E[FM] - E[A-AF] and E[C-AF] - E[G-AF], respectively, and for $J_a$
values of $-5$ and $+6$~meV  from  E[G-AF] - E[A-AF] and
E[FM] - E[C-AF], respectively. Clearly, the variation in the
values derived for $J_c$ ($J_a$) indicates that the changes
in the spin  configurations that we consider are too large to be
in the harmonic regime. These values, however, are nevertheless in 
order-of-magnitude  agreement with the experimental values of  $J_c =
17.5$~meV  and $J_a \approx -0.2$~meV.\cite{Hutchings69,Satija80}

\section{Influence of the magnetic structure on the orbital ordering 
and Jahn-Teller distortion}\label{sec:Influence}

\begin{figure}[t]
\includegraphics[width=8cm]{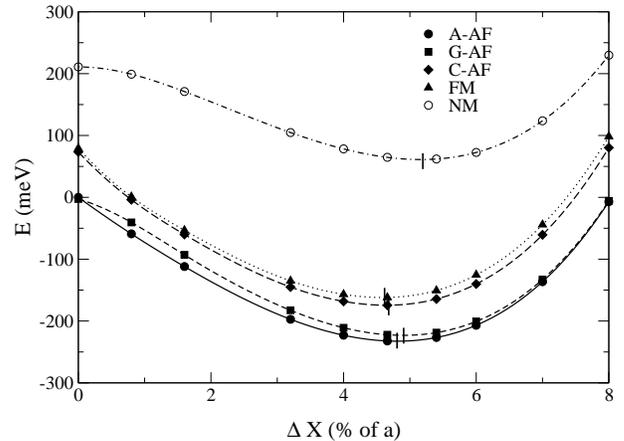}
\caption{\label{fig:Jahn_Teller}
Total energy of the various magnetic structures of KCuF$_3$
considered in this work as a function of the quadrupolar distortion
$\Delta X$.
 }
\end{figure}

In Fig.~\ref{fig:Jahn_Teller}, we display the total energy of  the
various spin structures considered in the previous section as a function 
of the quadrupolar distortion $\Delta X$. The calculated
equilibrium distortions range from 4.65\,\% (FM) to
5.15\,\% (NM), with an equilibrium value of 4.8\,\% for the A-AF
structure.  The  changes in the spin structure induce thus variations
in the Jahn-Teller distortion which do not exceed 10\,\%.
This 10\,\% value,  obtained when the  NM structure is
considered, is a  conservative upper bound for the estimated variation in
the Jahn-Teller distortion produced by the A-AF to paramagnetic
transformation. Indeed,
experimentally, strong antiferromagnetic spin correlations are known
to persist above the N\'eel temperature within the Cu chains along the 
$c$  axis. Locally, this  would be more realistically described using
configurations of the type G-AF and A-AF. For example, if one considers
a structure composed of Cu AF chains along the $c$ axis,
with no interchain magnetic interaction, and uses an Ising model to 
describe the spin interactions, the  energy of this structure is the
average of the energies of the G-AF and A-AF configurations.
Using such a description to model the structure above the N\'eel
temperature, and with the energies of the G-AF and A-AF configurations
given in Fig.~\ref{fig:Jahn_Teller}, the estimated change in the Jahn-Teller 
distortion produced by the magnetic transformation decreases to $\sim1$~\%.

To compare the orbital ordering in the different spin structures, we
have reported in  Table~\ref{tab:hole} the calculated
number of  $3d_{z^2-x^2}$ holes on the A site, $\bar{n}_{z^2-x^2}^{A,hole}$,
together with the total number of $3d$-$e_g$ holes per Cu site,
$n_{e_g}^{hole}$, in each structure.  These values have been calculated
using the same quadrupolar distortion for all structures, i.e., $\Delta X
= 4.6$~\%. If one uses, instead, the calculated equilibrium  $\Delta X $ for  each
structure,  the numbers, in Table~\ref{tab:hole}, change by less than 1\,\%.
For all spin structures, the orbital ordering is almost exclusively of the
$z^2-x^2$/$z^2-y^2$ type; the $d_{z^2-x^2}$/$d_{z^2-y^2}$ character, given by 
the ratio $n_{z^2-x^2}^{A,hole}/n_{e_g}^{hole}$, ranges from 91\,\% for the NM
structure to 98 \% for the G-AF structure.
The major change we observe  is a strong decrease ($\sim30$\,\%) in the
degree of orbital ordering, given by $\bar{n}_{z^2-x^2}^{A,hole} $,
 of the NM structure relative to those of the other
structures. This decrease is associated with an increased  delocalization of the
$3d$ hole. Although there is an important change thus  in the orbital ordering
---and $3d$ hole occupation---
induced by a change from the magnetic to the NM configuration, we will see 
in the next section that such a change has not a  major  impact on the  
RXS spectrum.

\begin{table}[t]
\begin{ruledtabular}
\begin{tabular}{lcc}
 & $ \bar{n}_{z^2-x^2}^{A,hole} $& $n_{e_g}^{hole}$ \\ \hline
A-AF &  0.61 & 0.63\\
G-AF &  0.62 & 0.63\\
C-AF &  0.62 & 0.66\\
FM   &  0.64 & 0.66\\
NM   &  0.42 & 0.46 \\
\end{tabular}
\end{ruledtabular}
\caption{\label{tab:hole} Calculated number of 3$d$-$e_g$ hole 
per Cu site, $n_{e_g}$,  and expectation value of the number of  
3$d_{z^2-x^2}$ hole on a Cu A site,
$\bar{n}_{d_{z^2-x^2}}^{A,hole}$, in the various spin structures 
of KCuF$_3$ considered in this work. 
}
\end{table}

\section{RXS spectra}\label{sec:RXS}

Resonant elastic x-ray scattering  is a second order process in which 
a core electron is virtually promoted to  some
intermediate states above the Fermi energy, and subsequently 
decays  to the same core level.  We concentrate here on
RXS near the Cu K-absorption edge, and examine the resonant 
scattering intensity  for  Bragg reflections that selectively probe 
the orbital order  (or equivalently the Jahn-Teller structural order).
The corresponding Bragg vectors are  $\bm{G} = (h ,k ,l )$, with  
$h,k,l$ odd,  in units  $ (\frac{\pi}{a}, \frac{\pi}{a},\frac{\pi}{c})$. 
We consider dipole transitions to Cu-$4p$ band states, and neglect 
the effect of the core-hole potential on the intermediate states.  

The RXS intensity for a Bragg  vector $\bm{Q}$ and
incoming- (outgoing-) photon energy $\hbar \omega$ 
may be written as:
\begin{equation}\label{eq:Idef}
I (\bm{G},\hbar \omega) \propto \left| \sum_j 
e^{i \bm{Q} \cdot \bm{R}_j }
\sum_{\alpha,\beta} F^j_{\alpha,\beta} (\hbar \omega)
{\epsilon}_{\alpha} {\epsilon'}_{\beta} \right|^2,
\end{equation}
where $\bm{R}_j $ are the positions of the Cu ions in
the unit cell, ${\epsilon}_{\alpha}$ $({\epsilon'}_{\beta})$
are the components of the polarization vector  of the
incoming (outgoing) photon,  $\alpha (\beta) = x,y,z$,  and
\begin{equation}\label{eq:Fdef}
F^j_{\alpha,\beta} (\hbar \omega) = \sum_{\bm{k},n}
\frac{ \langle \psi_0^{(j)} | r_{\alpha} (j)| \psi^{4p}_{\bm{k},n}    \rangle
\langle \psi^{4p}_{\bm{k},n} | r_{\beta}(j) |  \psi_0^{(j)} \rangle }
{\hbar \omega + E_0 - E^{4p}_{\bm{k},n} - i \Gamma/2} 
\end{equation}
are the components of the resonant atomic scattering tensor 
for the $j$th Cu ion in the cell.  
In Eq.~(\ref{eq:Fdef}),  $\psi_0^{(j)}$ is the Cu-1$s$  wave function
at site $j$,  $E_0$  is the corresponding core level energy,  
$ \bm{r}  (j)$ is the position operator measured relative to 
$\bm{R}_j $, $\psi^{4p}_{\bm{k},n}$ ($E^{4p}_{\bm{k},n}$) are the empty  
Cu-$4p$ band states (energies) with  Bloch vector $k$ and band index $n$, and 
$\Gamma$ is the broadening corresponding to the inverse lifetime
of the Cu $1s$-core-hole and $4p$-electron excited states.

For Bragg vectors associated with the orbital order, the structure 
factors, in  Eq.~(\ref{eq:Idef}), read:
\begin{equation}\label{eq:struc}
 \sum_j  e^{i \bm{Q} \cdot \bm{R}_j } F^j_{\alpha,\beta} 
= 2( F^A_{\alpha,\beta} - F^B_{\alpha,\beta}),  
\end{equation}
where $F^{A(B)}_{\alpha,\beta}$ is the atomic scattering tensor
of a Cu ion at site  A (B), and the factor of two  accounts 
for the two A (B) Cu ions in the unit cell. 
Because of the reflection symmetries $\sigma_x$, 
$\sigma_y$, and $\sigma_z$ present  at the Cu sites in KCuF$_3$, 
the only non-vanishing components of the atomic scattering tensors 
are: $F^{A(B)}_{x,x}$,  $F^{A(B)}_{y,y}$, and  
$F^{A(B)}_{z,z}$,  and because of the 90-degree roto-translation 
symmetry that transforms the A and B site into each other, one has:
$F^A_{x,x (y,y)} = F^B_{y,y (x,x)}$ and
$F^A_{z,z} = F^B_{z,z}$. The scattering intensity in 
Eq.~(\ref{eq:Idef}) may therefore be written as:
\begin{equation}\label{eq:I}
I (\bm{G},\hbar \omega) \propto \left|
F^A_{x,x} (\hbar \omega)-  F^A_{y,y}(\hbar \omega) \right|^2
\left| {\epsilon}_{x} {\epsilon'}_{x}  - {\epsilon}_{y} 
{\epsilon'}_{y} \right|^2.
\end{equation}

Except for the special case where  the polarization-dependent term
on the right-hand side of Eq.~(\ref{eq:I}) vanishes,   the scattering
intensity is  proportional thus to $ \left|  F^A_{x,x} (\hbar \omega)-
F^A_{y,y}(\hbar \omega) \right|^2  $.
We also note that,  within the extremely localized
core region where  the Cu-$1s$ wave function does not vanish, the
radial dependence of each band state  $\psi^{4p}_{\bm{k},n}$,
in Eq.~(\ref{eq:Fdef}), is essentially the same as that of the
atomic $4p$ state,  except for a global scaling factor in the
wave-function amplitude.
The scattering factors, in Eq.~(\ref{eq:I}), may therefore be written as:
\begin{equation}
F^A_{\alpha,\alpha} (E) \propto \int_{E_F}^{\infty} 
\frac{d\varepsilon D^{4p}_{\alpha}(\varepsilon)}
{E+E_0-\varepsilon - i \Gamma/2} ,
\end{equation}
where $D^{4p}_{\alpha}(\varepsilon)$ is the projected atomic
$4p_{\alpha}$ density of states and $E_F$ is the Fermi energy. 
Hence, in our pseudopotential calculations we evaluate the orbital 
scattering intensity,   
$ I_{\rm orb} (\hbar \omega) \propto \left|
F^A_{x,x} (\hbar \omega) -  F^A_{y,y}(\hbar \omega) \right|^2$, 
as a function of photon energy,
directly from the partial $4p$ densities of states. 
The core-level energy $E_0$ (not computed here) is a value 
we adjust to align the main feature of the calculated and  
experimental RXS spectrum.\cite{Note_excit} 

\begin{figure}[t]
\includegraphics[width=8cm]{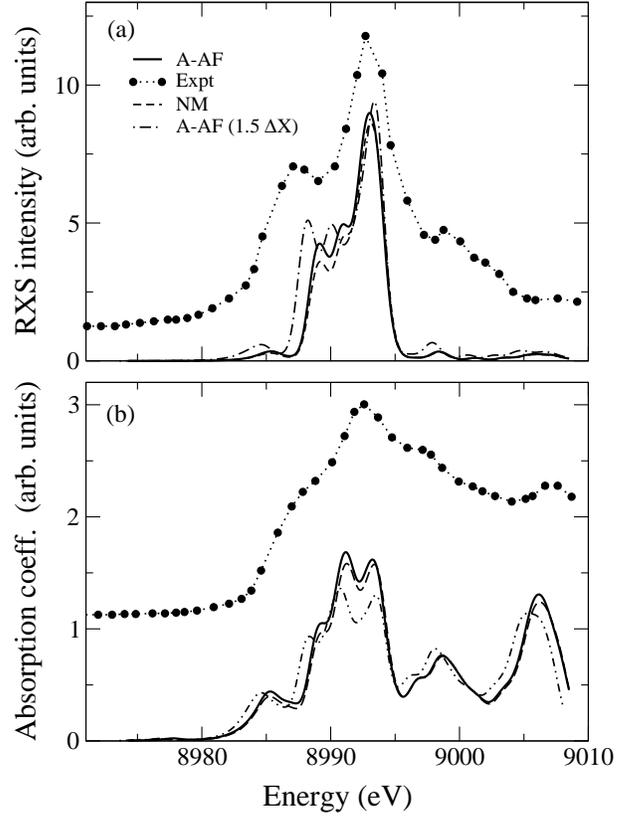}
\caption{\label{fig:RESX}
Orbital RXS intensity (a) and absorption coefficient (b), as a function of  photon energy,  
near the Cu K edge in KCuF$_3$. The calculated spectra of the A-AF (solid line) 
and NM (dashed line) structures are displayed. The effect of a 50\,\% increase 
in the quadrupolar distortion, $\Delta X$, in the A-AF structure
is also shown (dashed-dotted line). The experimental data are from
Ref.~\onlinecite{Paolasini02}; the measured RXS spectrum corresponds 
to a (3,3,1) orbital Bragg reflection and a $\sigma -\pi'$ polarization.
 }
\end{figure}

In Fig.~\ref{fig:RESX}(a), we confront our calculated orbital RXS 
spectrum for the A-AF structure of KCuF$_3$ to the experimental 
spectrum [measured at the (3,3,1) orbital Bragg reflection and 
for a $\sigma - \pi'$ polarization]. We also reported in
this figure the theoretical RXS spectrum of the NM structure, for
comparison. Both spectra have been computed using the experimental
values of the structural parameters. A spectrum obtained for the 
A-AF structure with a 50\% increase in $\Delta X$ is also shown
in Fig.~\ref{fig:RESX}(a). The corresponding calculated K-edge 
absorption spectra $A(\hbar \omega)$, evaluated as  $A(\hbar \omega) 
\propto  Im [F^A_{x,x} (\hbar \omega) + F^A_{y,y}(\hbar \omega) + 
F^A_{z,z}(\hbar \omega)] $, are displayed in 
Fig.~\ref{fig:RESX}(b), and also compared to experiment (fluorescence 
data).\cite{Paolasini02} 
All theoretical spectra have been evaluated using  for
$\Gamma$   0.5~eV,  and we convoluted these spectra with a  Gaussian 
of full-width at  half maximum of 1 eV.

The calculated RXS and absorption spectra for the A-AF structure are 
in reasonable agreement with the experimental data.
We note, however, that the energy separation between the two main 
features in the RXS spectrum is somewhat underestimated in our
calculations, and the intensity of the  high-energy feature near
9 keV is also not very well reproduced (both in the RXS and in the 
absorption spectrum).   
The change in the magnetic configuration, from the A-AF structure to the
NM structure,  decreases somewhat the RXS intensity. We note, however, 
that this   change is very  small compared
to the factor-of-two decrease in intensity observed experimentally when the
material transforms to the  paramagnetic phase. The weak dependence  of the  
RXS spectrum on the magnetic configuration is consistent with the
results of previous all-electron calculations for the A-AF structure
(calculated within the LDA+U) and the NM structure (calculated
within the LDA in previous studies).\cite{Caciuffo2002,Igarashi03}

To assess the  influence of the orbital ordering on the RXS
signal, we have also  computed the RXS intensity of the A-AF structure
without the quadrupolar distortion ($\Delta X\sim0$).
Although the $z^2-x^2$/$z^2-y^2$ orbital ordering is still large
in this case: $ \bar{n}_{z^2-x^2}^{A,hole}  =0.64$, $n_{e_g}^{hole} = 0.68$,
the calculated RXS signal is about  two orders of magnitude smaller
 than that obtained with the experimental Jahn-Teller distortion;
the orbital ordering by itself  thus has a negligible influence on the RXS
intensity. We   note that the fact that
$ \bar{n}_{z^2-x^2}^{A,hole}$ (and  $n_{e_g}^{hole}$) are very
similar with and without the quadrupolar distortion indicates that,  
already at vanishing $\Delta X$, the orbital ordering has essentially 
reached its saturation value, 
given the hybridization present in the system.

Concerning the effect of the Jahn-Teller distortion, we observe that 
increasing $\Delta X$ by 50\,\% increases noticeably the splitting
between the two mean features in the RXS spectrum of the A-AF structure, 
but does not change significantly the intensity of the dominant peak.
Decreasing, instead, $\Delta X$ by the same amount  has a larger influence
on the RXS intensity,\cite{Igarashi03}  but corresponds to   a situation where
the two features are essentially  merged.
A change in the Jahn-Teller distortion in the
magnetic phase (not supported by our LDA+U calculations of the
Jahn-Teller distortion)
has been suggested as a possible cause of the drastic
change observed experimentally in the  RXS signal near the N\'eel
temperature.\cite{Igarashi03}  
We observe here, however,  that a factor  of two change in
the RXS intensity would require a drastic change in $\Delta X$,
in the conventional crystallographic  structure of KCuF$_3$. 
Such a change is bound to induce also a substantial  change  in the 
$c/a$ ratio (see Section~\ref{sec:Lattice}),
which has  not been observed experimentally.
The Jahn-Teller origin of the jump in intensity appears thus somewhat unlikely.
We therefore suggest  that possibly a structural transformation
occurs at low temperature in KCuF$_3$,  which is reflected in the RXS signal. 
We note that such a change in the crystal structure would not be inconsistent
with some unexplained Raman\cite{Ueda91} and nuclear magnetic 
resonance\cite{Mazzoli02} features observed experimentally  at low temperature
in this material.
We also note that recently a new crystal structure (superstructure)
has been proposed for KCuF$_3$ based on x-ray diffraction at room 
temperature.\cite{Hidaka98}
In our LDA+U calculations, however,  his structure  is found to relax to the
conventional  crystallographic structure of KCuF$_3$.

\section{Conclusions}\label{sec:Summary}

We have studied  by means of  LDA+U pseudopotential calculations,
the structural, electronic, and magnetic properties  of KCuF$_3$,
and  investigated  the Cu K-edge RXS spectrum  for Bragg reflections
associated with orbital order.
In our studies, we considered several different
(meta)-stable spin structures for KCuF$_3$ in order to assess  the influence
of the magnetic structure on the orbital ordering, Jahn-Teller distortion, and
RXS spectrum.

For KCuF$_3$, the LDA fails to correctly predict the
stable AF insulating structure.
We have found here that LDA+U pseudopotential calculations provide  an
accurate general  description of the properties of this system.
The ground state is correctly  predicted to be an A-type AF
structure. The structural parameters obtained from the LDA+U calculations 
agree also well with experiment, with an accuracy comparable to that obtained
by LDA calculations for other perovskite materials. 
The orbital ordering is predominantly of the $z^2-x^2$/$z^2-y^2$
type, consistent with previous theoretical predictions for this system. 
We  find, however, that  the orbital-order parameter is significantly
reduced with respect to its nominal value due to Cu($3d$)--F($2p$) 
hybridization.  We also find that, given this hybridization, the  
Cu-$3d$ orbital ordering in the A-AF phase is already saturated at vanishing 
Jahn-Teller quadrupolar distortion of the F neighbours. 

The  RXS spectrum calculated for the A-AF  structure  agrees  relatively
well with the experimental spectrum. 
We find that the  resonant signal is dominated by the Jahn-Teller distortion,
with   a minor influence of the orbital ordering, in
agreement with previous theoretical results.
Our  LDA+U calculations, however,  also indicate that a change in the 
magnetic  structure has a small  influence  on the Jahn-Teller
distortion, and hence on the resonant  spectrum,  in the commonly
accepted crystallographic structure of KCuF$_3$.
We therefore suggest that the
change observed experimentally  in the RXS signal
near the N\'eel temperature may be related to  a low-temperature structural 
transformation  in KCuF$_3$.  
We also note that preliminary results obtained from LDA+U molecular-dynamic simulations
for A-AF KCuF$_3$ indicate the existence of a  superstructure, with tilted CuF$_6$
octahedra, which is slightly lower in energy than the conventional crystal structure
of  KCuF$_3$.

\begin{acknowledgments}

It is a pleasure to thank R. Caciuffo, S. de Gironcoli, M. Cococcioni,
G. Trimarchi, and N. Stojic for useful discussions. The calculations 
are based on the PWSCF code.\cite{PWSCF}
We acknowledge support for this work by the INFM  within the 
framework of the ``Iniziativa Trasversale di Calcolo Parallelo''.

\end{acknowledgments}


\begin{thebibliography}{99}
\bibitem{Note_proj}  A tight-binding Bloch state, normalized over 
the crystal unit cell, is generated for each localized atomic  
function $\varphi^{I,\sigma}_m(\bm{r})$ and used to perform the 
projection.  We note that with the ultrasoft-pseudopotential scheme, the
projection includes also the augmentation pseudo-charge density (see 
Ref.~\onlinecite{Cococcioni}).
\bibitem{Note_pdos} The projected atomic DOS's are evaluated from projections
within a small atomic-core sphere: 
$\int d\Omega \int_0^{\rm{rmax}} dr r^2 \varphi^{*}_m (\bm{r})
 \psi_{\bm{k},i} (\bm{r}),$ where $\rm{rmax} = 0.5$~a.u.. Changing
 $\rm{rmax}$ to 1 a.u. or 0.2 a.u. leaves the projected DOS's unchanged
 (except for a constant DOS scaling factor).
\end{thebibliography}
\end{document}